\RequirePackage{ifpdf}
\documentclass[11pt,a4paper,notoc]{article}
\usepackage{jheppubme}

\font\cmss=cmss12

\newcommand\half{\frac12}

\newcommand\bi{\begin{itemize}}
\newcommand\ei{\end{itemize}}

\newcommand\bea{\begin{eqnarray}}
\newcommand\eea{\end{eqnarray}}
\newcommand\be{\begin{equation}}
\newcommand\ee{\end{equation}}

\newcommand\ZZ{\hbox{Z\kern-.4emZ}}
\newcommand\sZZ{\hbox{\sevenfont Z\kern-.4emZ}}

\newcommand{\eref}[1]{Eq.\,(\ref{#1})}
\newcommand{\Comment}[1]{{}}

\def\IB{\relax{\rm I\kern-.18em B}}
\def\IC{{\relax\hbox{\kern.3em{\cmss I}$\kern-.4em{\rm C}$}}}
\def\ID{\relax{\rm I\kern-.18em D}}
\def\IE{\relax{\rm I\kern-.18em E}}
\def\IF{\relax{\rm I\kern-.18em F}}
\def\II{\relax{\rm I\kern-.18em I}}
\def\IZ{\mathbb{Z}}
\def\Id{\relax{1\kern-.32em 1}}
\def\IG{\relax\hbox{$\inbar\kern-.3em{\rm G}$}}
\def\IR{\relax{\rm I\kern-.18em R}}

\title{Classification of RCFT from Holomorphic Modular Bootstrap: A Status Report\footnote{To appear in the Proceedings of the Pollica Summer Workshop on ``Mathematical and Geometric Tools for Conformal Field Theories'', June 3-21, 2019.}} \author{Sunil Mukhi\,\footnote{Email:
    sunil.mukhi@gmail.com}\\ \it Indian Institute of Science Education and Research,\\ \it Homi Bhabha Rd, Pashan, Pune 411 008, India} 

\abstract{Following the initial proposal in 1988, there has been much progress in classifying Rational Conformal Field Theories in 2 dimensions from the Holomorphic Bootstrap approach. This method starts by postulating a generic holomorphic Modular Linear Differential Equation of a given order and imposing the requirement of non-negative integrality of the coefficients in the series expansion of the solutions, which are then identified as admissible characters, from which a modular-invariant partition function is constructed. In this short note, the status of this project is summarised.}

\preprint{}


\notoc

\begin{document}

\maketitle

\section{Introduction}

\label{intro}

Conformal invariance in two dimensions has long been recognised as being a strong enough symmetry to partially classify, and solve, field theories having this invariance. An influential paradigm, initiated in \cite{Belavin:1984vu, Knizhnik:1984nr} has been to focus on the chiral symmetry algebra, identify its null vectors and search for the ``minimal'' models that arise on decoupling them.  These lead to infinite families of conformal field theories with rational exponents. The Kac-Moody algebra is special because one can also take cosets of pairs of such models and obtain enormous families of new ones \cite{Goddard:1984vk, Goddard:1986ee}. Within each family (labelled by central charge or Kac-Moody level) the number of primaries under the chiral algebra increases rapidly down the list, and only the first few models are simple enough to be interesting.

The holomorphic bootstrap takes a different approach and gives rise to a classification based on the number of generalised characters. It is based on the fact that the characters $\chi_i(\tau),~i-0,1,\cdots n-1$ of an RCFT are holomorphic functions of the modular parameter $\tau$ and transform as vector-valued modular functions of weight zero:
\be
\chi_i(\gamma\tau)=V_{ij}(\tau)\chi_j(\tau),\qquad \gamma\tau\equiv \frac{a\tau+b}{c\tau+d},~ \begin{pmatrix} a &b\\ c&d\end{pmatrix}\in \hbox{PSL(2,Z)}
\label{modvec}
\ee
which ensures that the partition function:
\be
Z(\tau,{\bar\tau})=\sum_{i=0}^{n-1}{\bar\chi}_i({\bar\tau})\chi_i(\tau)
\ee
is modular invariant. 

\eref{modvec} holds if and only if $\chi_i(\tau)$ are the $n$ independent solutions of a Modular Linear Differential Equation (MLDE) \cite{Mathur:1988rx,Mathur:1988na,Mathur:1989pk}. As proposed in these references, one therefore starts with a generic MLDE -- which already guarantees that its solutions have the desired holomorphic and modular properties. The equation turns out to depend on finitely many real parameters. Now one solves the equation by inserting a series expansion in $q=e^{2\pi i\tau}$ and  imposes the requirement that the coefficients in this expansion are non-negative integers. This is necessary in order to interpret them as degeneracies in a quantum field theory. The reader is referred to the original references for details on the procedure.

What will be important for us is that the MLDE is specified by two integers: $n$, the number of characters, and $\ell$, the Wronskian index which is equal to the number of zeroes in moduli space of the Wronskian determinant of the solutions. For fixed $(n,\ell)$, the outcome of the bootstrap is to produce lists of ``admissible'' characters satisfying all the bootstrap conditions. A further problem, after such a list is generated, is to identify which candidates can be identified as the characters of actual conformal field theories. In the present note I will focus primarily on the classification of admissible characters, and refer to this additional question only in a few places.

\section{Summary of results}

\subsection{$n=1$}

For single-character CFT, candidate characters take the form \cite{Mathur:1988na}:
\be
\chi(\tau)=j^{w_\rho}(j-1728)^{w_i} P_{w_\tau}(j)
\label{onechar}
\ee
where $j(\tau)$ is the Klein $j$-invariant, $w_\rho\in \{0,\frac13,\frac23\}$, $w_i\in \{0,\half\},w_\tau\in\IZ$ and $P_{w_\tau}(j)$ is a polynomial of degree $w_\tau$ in $j$. The polynomial must be chosen so that the final character is admissible, i.e. has non-negative coefficients. The Wronskian index for these characters is:
\be
\ell=6(w_\rho+w_i+w_\tau)
\ee
and the central charge of the corresponding theory, if any, is $c=4\ell$. This set is finite for $\ell<6$ ($c<24$) and infinite beyond that. Corresponding CFT are completely classified for $\ell <6$ while seminal work was done on the case of $\ell=6$, equivalently $c=24$ \cite{Schellekens:1992db}.

\subsection{$n=2, \ell<6$}

\label{sec.twochar}

For $\ell=0$, a list of ten pairs of admissible characters was found in the original work, \cite{Mathur:1988na}. Of these, one corresponds to a one-character theory re-discovered as a solution to a second-order MLDE, namely the WZW  model $E_{8,1}$. Another corresponds to the Lee-Yang edge singularity, a non-unitary minimal CFT. Of the remaining cases, seven correspond to WZW models for Lie algebras in the ``miraculous'' Cvitanovic-Deligne series \cite{Cvitanovic:2008zz, Deligne:1}. One corresponds to the so-called $E_{7\half}$ algebra \cite{Landsberg:2004} and has been identified, together with the Lee-Yang theory, as an ``intermediate vertex operator algebra'' in \cite{Kawasetsu:2014}. The central charges of this series satisfy $0<c\le 8$ in the ``unitary presentation'' (see \cite{Chandra:2018pjq}). A mathematically rigorous basis for this classification was provided in \cite{Mason:2018}.

Subsequently it was shown in \cite{Naculich:1988xv} that for $n=2$, the Wronskian index $\ell$ is always even. The case of $(n,\ell)=(2,2)$ was classified in the same reference. There are again 10 cases, much like the $\ell=0$ set. Their central charges satisfy $16\le c < 24$. Many years later these were re-examined in \cite{Hampapura:2015cea}. In \cite{Gaberdiel:2016zke} it was shown that the $\ell=0$ and $\ell=2$ sets form dual coset pairs with their central charges adding up to 24. This novel coset procedure is quite general and will re-appear below.

For $(n,\ell)=(2,4)$ the situation remains puzzling. After being missed in previous works, two of the admissible characters were found in \cite{Tener:2016lcn} and a third was added in \cite{Chandra:2018pjq}. Their central charges lie between 32 and 35, but no CFT's corresponding to these characters have yet been identified. Besides these, there are tensor products of $\ell=0$ characters with the $E_{8,1}$ character $j^\frac{1}{3}$.

\subsection{$n=2,\ell \ge6$}

The case of $\ell\ge 6$ was addressed by different methods. In \cite{Harvey:2018rdc} such characters were constructed as ``Hecke images'' of characters with lower $\ell$. It became clear from this reference that there are infinitely many candidate characters in this case. In \cite{Chandra:2018pjq} it was shown that admissible characters can be constructed as semi-definite linear combinations of $\ell=0$ ``quasi-characters'', which themselves have integer but possibly negative coefficients in their $q$-series. The latter reference classified all quasi-characters and showed that by suitable linear combinations, one generates the complete set of characters with $\ell\ge 6$.

The admissible characters so obtained are infinite in number and do not in general have any recognisable CFT interpretation. From experience with the one-character case, it is clear that once we cross the ``$\ell=6$ barrier'' there should be far fewer CFT's than admissible characters. In \cite{Chandra:2018ezv}, over a hundred genuine RCFT were constructed with $\ell=6$ using the coset construction of \cite{Gaberdiel:2016zke} in conjunction with Kervaire lattice CFT. These have central charges satisfying $24<c < 32$ and provide an existence proof for RCFT with $(n,\ell)=(2,6)$.

\subsection{$n=3,\ell=0$}

For three characters the only well-studied case is $\ell=0$. This was first addressed in \cite{Mathur:1988gt} and discussed further via a different technique in \cite{Mukhi:1989qk}. It was noted that with three characters, even at $\ell=0$ there are infinitely many known RCFT, and several examples as well as some bounds were obtained. In particular, the WZW model for SO$(N)_1$, SU$(4)_1$, SU$(5)_1$, SU$(2)_2$ as well as the Virasoro minimal models with $(p,p')=(3,4)$ (Ising) and $(2,7)$ (a non-unitary theory) all have three characters as well as $\ell=0$.

Coset duals of many three-character $\ell=0$ theories were obtained in \cite{Gaberdiel:2016zke}. These also turn out to have $\ell=0$ and are listed in Table 2 of that reference. Also, all three-character theories with $\ell=0$ and no Kac-Moody algebra were classified in \cite{Tuite:2008pt, Hampapura:2016mmz}. This turns out to be a finite set containing the Baby Monster CFT with $c=\frac{47}{2}$ \cite{Hoehn:Baby8} as well as exotic theories at $c=8,16$ related to the finite group $O_{10}^+(2)$. These CFT's were rediscovered in \cite{Bae:2018qym} in the context of the semi-definite programming approach to the conformal bootstrap.

Recently in \cite{Franc:2019},  with the extra condition of ``irreducible monodromy'', a complete classification of three-character theories with $\ell=0$ has been proposed. The results are the examples discussed in \cite{Mathur:1988gt} along with a ``$U$-series'' (where $U$ stands for ``unknown'') which includes the Baby Monster and a set of other theories with central charges in the range $\frac{27}{2},\frac{29}{2},\cdots \frac{47}{2}$. 
Many, but not all, of these $U$-series theories have previously appeared in 
Table 2 of \cite{Gaberdiel:2016zke}, namely those with central charges:
\be
c=\frac{31}{2},\frac{35}{2},\frac{37}{2},\frac{39}{2},\frac{41}{2},\frac{43}{2},\frac{45}{2}
\ee

There is a subtlety for the sub-series SO$(2N)_1$ for $N$=0 mod 4. Though these certainly exist as RCFT, they do not satisfy the irreducible monodromy condition of \cite{Franc:2019} (which is violated if two distinct primaries differ in dimension by an integer). This subtlety appears to be associated to congruence subgroups, which play an important role in the classification procedure of this work. The exotic $c=8,16$ models of \cite{Tuite:2008pt, Hampapura:2016mmz,Bae:2018qym} also do not satisfy the irreducible monodromy condition and therefore do not appear in this classification, though they appear to be sensible RCFT. Moreover, there are several other theories in Table 2 of \cite{Gaberdiel:2016zke}, with central charges: 
\be
c=14,15,17,18,19,20,21
\ee
that do not appear in \cite{Franc:2019}. None of these has a pair of primaries differing in dimension by an integer, so naively it seems they should satisfy the criterion of irreducible monodromy. It will be interesting to establish why they do not appear in the classification of \cite{Franc:2019}. 

It seems quite a realistic goal to complete the classification of admissible characters for $(n,\ell)=(3,0)$, including cases with reducible monodromy, and even to associate most/all of them to actual RCFT. However little or nothing seems to be known about the case of $\ell>0$. 

\subsection{$n=4$}

Very few results are available for this case. To my knowledge, the only ones (beyond the usual set of minimal and WZW models) are the $\ell=0$ coset models obtained in Table 3 of \cite{Gaberdiel:2016zke}. There is a paper in the mathematical literature \cite{Arike:2018fru} where the 4th order MLDE is solved under very restrictive conditions on the degeneracies of the identity character. Four RCFT are obtained, all of which are either minimal models or simple-current extensions.

\section*{Acknowledgements}

I am grateful to the organisers of the Pollica Summer Workshop,  which was supported in part by the Simons Foundation (Simons Collaboration on the Non-perturbative Bootstrap) and in part by the INFN,  for providing a splendid setting and a wonderful academic atmosphere in which to exchange scientific ideas. A grant from Precision Wires Ltd., India, made this trip possible and is gratefully acknowledged.

\bibliographystyle{JHEP}

\bibliography{Pollica}

\end{document}